\documentclass[twocolumn,showpacs,preprintnumbers,amsmath,amssymb]{revtex4-1}

\usepackage{graphicx}
\usepackage{dcolumn}
\usepackage{bm}
\usepackage{color}

\begin{document}

\preprint{}

\title{Conduction electrons localized by charged magneto-acceptors {\itshape A}$^{2-}$ in GaAs/GaAlAs quantum wells}

\author{M. Kubisa}
\email{Maciej.Kubisa@pwr.edu.pl}
\author{K. Ryczko}
\affiliation{Laboratory for Optical Spectroscopy of Nanostructures, 
Department of Experimental Physics, Wroc{\l}aw University of Technology, Wybrze{\.z}e Wyspia{\'n}skiego 27, 50-370 Wroc{\l}aw, Poland}
\author{I. Bisotto}
\affiliation{LNCMI, UPR 3228, CNRS-INSA-UJF-UPS, BP166, 38042 Grenoble, Cedex 9, France}%
\author{C. Chaubet}
\author{A. Raymond}
\affiliation{L2C UMR 5221, CNRS-Université Montpellier 2, Place E. Bataillon, 34090 Montpellier cedex 05, France}%
\author{W. Zawadzki}
\affiliation{Institute of Physics, Polish Academy of Sciences, 02668 Warsaw, Poland}%

\date{\today}
						
\begin{abstract}
A variational theory is presented of {\itshape A}$^{1-}$ and {\itshape A}$^{2-}$ centers, i.e. of a negative acceptor ion localizing one and two conduction electrons, respectively, in a GaAs/GaAlAs quantum well in the presence of a magnetic field parallel to the growth direction. A combined effect of the well and magnetic field confines conduction electrons to the proximity of the ion, resulting in discrete repulsive energies above the corresponding Landau levels. The theory is motivated by our experimental magneto-transport results which indicate that, in a heterostructure doped in the GaAs well with Be acceptors, one observes a boil-off effect in which the conduction electrons in the crossed-field configuration are pushed by the Hall electric field from the delocalized Landau states to the localized acceptor states and cease to conduct. A detailed analysis of the transport data shows that, at high magnetic fields, there are almost no conducting electrons left in the sample. It is concluded that one negative acceptor ion localizes up to four conduction electrons.
\end{abstract}

\maketitle

\section{INTRODUCTION} 
It is well known that, in bulk semiconductors, donors produce bound electron states below the bottom of the conduction band, while acceptors produce bound electron states above the top of the valence band. This is related to the sign of potential energy, negative for donor ions and positive for acceptor ions. A conduction electron is attracted to the positive charge of the donor ion and repulsed by the negative charge of the acceptor ion. Thus, ionized acceptors participate in the scattering of conduction electrons but do not form bound electron states in the conduction band. In two-dimensional (2D) structures the situation may be different. Let us consider a GaAs/GaAlAs heterojunction doped in the GaAlAs barrier with donors and in the GaAs well with a considerably smaller density of acceptors. Electrons from the donors go into the well and ionize the acceptors. An ionizing electron is near the acceptor nucleus since the Bohr radius is small, being determined by the heavy-hole mass. As a consequence, one deals with negative acceptor ions that interact in the well with remaining 2D conduction electrons. It is known that, in the presence of a magnetic field parallel to the growth direction, negative potential fluctuations broaden the Landau levels (LLs) on the lower energy sides, whereas positive potential fluctuations broaden LLs on the higher energy sides. Kubisa and Zawadzki~\cite{kubisa_1,kubisa_2} went a step further observing that, in this situation, a combined effect of quantum well and magnetic field keeps the conduction electron in the proximity of negative repulsive acceptor ion, forming a system which we call {\itshape A}$^{1-}$.  More specifically, the electron cannot run away from the negative acceptor ion along the growth direction \textit{z} because of the well and along the \textit{x-y} plane because of the Lorentz force induced by the magnetic field that keeps it on the cyclotron orbit around the ion. It was shown with the use of variational calculations that this confinement results in {\itshape discrete repulsive energies} of the conduction electrons above the corresponding LLs, see also Refs.~\cite{laughlin} and~\cite{yang}. The discrete energies of {\itshape A}$^{1-}$ centers in the conduction band of acceptor-doped heterostructures were observed experimentally, first in photo-magneto-luminescence~\cite{vincente}, then in cyclotron resonance~\cite{bonifacie}, and finally in quantum magneto-transport~\cite{bisotto}. In the quantum transport, the discrete states of {\itshape A}$^{1-}$ centers are manifested by the so called rain-down and boil-off effects in which the electrons fall down from the localized {\itshape A}$^{1-}$ states to the delocalized Landau states (rain-down effect) or are transferred back at higher electric fields from the Landau states to the {\itshape A}$^{1-}$ states (boil-off effect).

When performing the quantum-transport experiments on Be-doped GaAs/GaAlAs heterostructures it was found that, in sufficiently high Hall fields, one can reach a situation in which a negative ion can localize two or more conduction electrons. Motivated by the above results we undertake in the present work a theoretical and experimental analysis of this situation. More specifically, we develop a theory of two conduction electrons kept in the proximity of a negative acceptor ion by a combined effect of the quantum well and external magnetic field, calling such a system {\itshape A}$^{2-}$center. This theory presents a generalization of the theory for {\itshape A}$^{1-}$ centers given in Refs.~\cite{kubisa_1,kubisa_2}. Further, we provide an experimental evidence that, in the Hall configuration at high Hall fields, there remain in the acceptor-doped heterostructures almost no conducting electrons and explain this phenomenon.

The problem of {\itshape A}$^{2-}$ center is reminiscent of a positive donor ion binding two conduction electrons in the presence of a magnetic field, which was called {\itshape D}$^{-}$ center. The {\itshape D}$^{-}$ centers were extensively investigated theoretically and experimentally in three-dimensional and two-dimensional systems, see for example Refs.~\cite{avron, larsen_3, huant, pang, dzyubenko}. However, it turns out that, in spite of an apparent similarity of the two situations, the physics of a negative acceptor ion with two conduction electrons is quite different from that of the {\itshape D}$^{-}$ center.

Our paper is organized as follows. In Section II we present the theory of a negative screened acceptor ion localizing one and two conduction electrons and give results for the corresponding repulsive energies. In Section III we report and analyze experimental magneto-transport data obtained on acceptor-doped GaAs/GaAlAs heterostructures at high magnetic fields. Section IV contains discussion of the results, the paper is concluded by a summary.

\section{THEORY} 
In the following section we describe a possibility to localize two conduction electrons on a negative acceptor ion in a heterojunction in the presence of an external magnetic field. This theory generalizes the description of a single electron in the same conditions, as described in Refs.~\cite{kubisa_1, kubisa_2}. We do not aim here at a high precision of calculated energies. Our purpose is to show that, since it is a combined effect of potential well and magnetic field that keeps the electrons in the proximity of a negative ion, the number of conduction electrons is not essential and one obtains discrete repulsive energies above the conduction Landau levels also for more than one electron. 

We consider a pair of conduction electrons at positions {\textit{\textbf r}}$_j=(x_j, y_j, z_j)=({\boldsymbol \rho_{j}}, z_j)$, where $j=1, 2$, in a heterojunction described by the potential $U(z)$. The electrons move in the presence of a magnetic field \textit{\textbf{B}}$||z$ and interact with an ionized acceptor located at \textit{\textbf{r}}$_{o}=(0, 0, z_o)$. The initial Hamiltonian for the problem reads
\begin{eqnarray}
H = && \sum_{j=1,2}\left[\frac{1}{2m^{*}}({\boldsymbol p_j}+\frac{e}{c}{\boldsymbol A_j})^2 + \frac{e^2}{\epsilon|{\boldsymbol r_j}-{\boldsymbol r_{o}}|} + U(z_j)\right] \nonumber\\ 
&& + \frac{e^2}{\epsilon|{\boldsymbol r_1}-{\boldsymbol r_2}|},
\label{eq:1}
\end{eqnarray}
where $m^{*}$ is the effective mass, $\epsilon$ is the dielectric constant, and {\textit {\textbf A}}$_j=[-\frac{1}{2}By_j, \frac{1}{2}Bx_j, 0]$ is the vector potential of  the magnetic field. Using the center of mass coordinate \textit{\textbf{R}} and relative coordinate \textit{\textbf{r}} of the electron pair in the {\itshape x-y} plane
\begin{subequations}
\label{eq:2}
\begin{equation}
{\textbf R}=\frac{\boldsymbol{\rho_{1}}+\boldsymbol{\rho_{2}}}{2}=[R\cos\Phi, R\sin\Phi],
\label{subeq:21}
\end{equation}
\begin{equation}
{\textbf r}=\boldsymbol{\rho_{1}}-\boldsymbol{\rho_{2}}=[r\cos\phi, r\sin\phi],  
\label{subeq:22}
\end{equation}
\end{subequations} 
and expressing the energy in effective Rydbergs $Ry^{*}=m^{*}e^{4}/2\epsilon^{2}\hbar^{2}$ and lengths in the effective Bohr radii $a_{B}^{*}=\epsilon\hbar^{2}/m^{*}e^{2}$, one obtains
\begin{eqnarray}
H = && -\frac{1}{2R}\frac{\partial}{\partial R}(R\frac{\partial}{\partial R})
-\frac{1}{2R^2}\frac{\partial^2}{\partial \Phi^2}-i\gamma
\frac{\partial}{\partial \Phi}+\frac{\gamma^2R^2}{2} \nonumber\\ 
&& -\frac{2}{r}\frac{\partial}{\partial r}(r\frac{\partial}{\partial r}) - \frac{2}{r^2}\frac{\partial^2}{\partial \phi^2} - i\gamma\frac{\partial}{\partial \phi}+\frac{\gamma^2r^2}{8} \nonumber\\
&& +\frac{2}{\sqrt{R^2+r^2/4+Rrcos(\Phi-\phi)+(z_1-z_o)^2}} \nonumber\\
&& +\frac{2}{\sqrt{R^2+r^2/4-Rrcos(\Phi-\phi)+(z_2-z_o)^2}} \nonumber\\
&& +\frac{2}{\sqrt{r^2+(z_1-z_2)^2}}-\frac{\partial^2}{\partial z_1^2}+U(z_1)
\nonumber\\ 
&&-\frac{\partial^2}{\partial z_2^2}+U(z_2).
\label{eq:3}
\end{eqnarray}
Here $\gamma=\hbar\omega_c/2Ry^{*}$, where $\omega_c=eB/m^{*}c$ is the cyclotron frequency. We reduce our 3D problem to an effective 2D problem~\cite{brum, kubisa_2} by seeking the wave function in the form of a product
\begin{equation}
	\Psi({\boldsymbol R},{\boldsymbol r},z_{1},z_{2})=F({\boldsymbol R},{\boldsymbol r})f_{o}(z_{1})f_{o}(z_{2}).
\end{equation}
The function $f_{o}(z)$ is the normalized eigenfunction of the electron in the lowest electric subband. It  satisfies the equation
\begin{equation}
	-\frac{d^{2}f_{o}(z)}{dz^2}+U(z)f_{o}(z)=E_{o}f_{o}(z),
\end{equation}
where $E_{o}$ is the energy of subband edge. By multiplying the eigenenergy equation $H\Psi=E\Psi$ by the product 
$f_{o}(z_{1})f_{o}(z_{2})$ and integrating over $z_{1}$ and $z_{2}$, one obtains the equation for \textit{F({\textbf R},{\textbf r})} in the form: $H_{2D}F=(E-2E_{o})F$, where
\begin{subequations}
\label{eq:6}
\begin{equation}
H_{2D}=H_{R}+H_{r}+V_{1}+V_{2}+V_{12}   
\label{subeq:61}
\end{equation}
is the effective 2D Hamiltonian for the pair of electrons. Here
\begin{equation}
H_{R} = -\frac{1}{2R}\frac{\partial}{\partial R}(R\frac{\partial}{\partial R})
-\frac{1}{2R^2}\frac{\partial^2}{\partial \Phi^2}-i\gamma
\frac{\partial}{\partial \Phi}+\frac{\gamma^2R^2}{2}  
\label{subeq:62}
\end{equation}
and
\begin{equation}
H_{r} = -\frac{2}{r}\frac{\partial}{\partial r}(r\frac{\partial}{\partial r}) - \frac{2}{r^2}\frac{\partial^2}{\partial \phi^2} - i\gamma\frac{\partial}{\partial \phi}+\frac{\gamma^2r^2}{8}  
\label{subeq:63}
\end{equation}
represent the kinetic energies of the center of mass and relative motions, 
\begin{equation}
V_{1}=\int^{\infty}_{-\infty}\frac{2f_{o}^{2}(z_{1})dz_{1}}{\sqrt{R^2+r^2/4+Rr\cos(\Phi-\phi)+(z_1-z_o)^2}}
\label{subeq:64}
\end{equation}
and
\begin{equation}
V_{2}=\int^{\infty}_{-\infty}\frac{2f_{o}^{2}(z_{2})dz_{2}}{\sqrt{R^2+r^2/4-Rr\cos(\Phi-\phi)+(z_2-z_o)^2}} 
\label{subeq:65}
\end{equation}
are the effective 2D impurity potentials, and finally 
\begin{equation}
V_{12}=\int^{\infty}_{-\infty}dz_{1}\int^{\infty}_{-\infty}dz_{2}\frac{2f_{o}^{2}(z_{1})f_{o}^{2}(z_{2})}{\sqrt{r^2+(z_{1}-z_{o})^2}}  
\label{subeq:66}
\end{equation}
\end{subequations}
is the effective 2D potential of electron-electron interaction.

Energy $E_{2e}$ of the ground state of the Hamiltonian (6) is evaluated by the variational method. We use the following two-parameter trial function
\begin{equation}
	F_{2e}(r,R)=\frac{1}{2\pi\alpha\beta}\exp(-\frac{R^2}{2\alpha^2}-\frac{r^2}{8\beta^2}).
\end{equation}
The variational parameters $\alpha$ and $\beta$ have simple physical meanings: $\alpha$ is the radius of the center-of-mass motion around the acceptor, while $\beta$ is the average distance between two electrons. Since the electrons are indistinguishable, the total wave function must change sign under the permutation of particles. To satisfy this condition, the ground state (7) corresponds to the singlet state having electron spins in opposite directions. Our trial function is less refined than that of the Chandrasekhar type used in the studies of $D^{-}$ centers~\cite{larsen_1}. As we mentioned above, our main goal is to show that a negative charged acceptor can localize a pair of 2D electrons. This does not require sophisticated trial functions.

The expectation value of {\itshape H}$_{2D}$ can be evaluated as
\begin{widetext}
\begin{subequations}
\label{eq:8}
\begin{equation}
\left\langle F_{2e}|H_{R}|F_{2e} \right\rangle = \frac{\gamma^2\alpha^2}{2} + \frac{1}{2\alpha^2},
\label{subeq:81}
\end{equation}
\begin{equation}
\left\langle F_{2e}|H_{r}|F_{2e} \right\rangle = \frac{1}{2\beta^2} + \frac{1}{2}\gamma^2\beta^2, 
\label{subeq:82}
\end{equation}
\begin{equation}
\left\langle F_{2e}|V_{1}|F_{2e} \right\rangle = \left\langle F_{2e}|V_{2}|F_{2e} \right\rangle = 2\int^{\infty}_{0} R(q,z_o)L_{o}(\frac{\beta^2q^2}{4})\exp(-\frac{\alpha^2+\beta^2}{4}q^2)dq, 
\label{subeq:83}
\end{equation}
\begin{equation}
\left\langle F_{2e}|V_{12}|F_{2e} \right\rangle = 2\int^{\infty}_{0}H(q)L_{o}(\beta^2q^2)\exp(-\beta^2q^2)dq, 
\label{subeq:84}
\end{equation}
\end{subequations}
\end{widetext}
where
\begin{subequations}
\label{eq:9}
\begin{equation}
R(q,z)=\int^{\infty}_{-\infty}f_{o}^{2}(z^{'})exp(-q|z'-z|)dz', 
\label{subeq:91}
\end{equation}
\begin{equation}
H(q)=\int^{\infty}_{-\infty}f_{o}^{2}(z)R(q,z)dz, 
\label{subeq:92}
\end{equation}
\end{subequations}
and $L_{o}(x)$ is the Laguerre polynomial. The formulas presented above apply to the case of unscreened Coulomb potential. It can be shown that, in order to account for the screening by 2D electrons in the manner described in Ref.~\cite{price} the expectation values of Coulomb terms in Eqs.(8) should be replaced by
\begin{widetext}
\begin{subequations}
\label{eq:10}
\begin{equation}
\left\langle F_{2e}|V_{1}|F_{2e} \right\rangle = \left\langle F_{2e}|V_{2}|F_{2e} \right\rangle = 2\int^{\infty}_{0}\frac{R(q,z_{o}}{1-\Pi(q)H(q)}L_{o}(\frac{\beta^2q^2}{4})\exp(-\frac{\alpha^2+\beta^2}{4}q^2)dq, 
\label{subeq:101}
\end{equation}
\begin{equation}
\left\langle F_{2e}|V_{12}|F_{2e} \right\rangle = 2\int^{\infty}_{0}\left[H(q)+\frac{R(q,0)\Pi(q)G(q)}{1-\Pi(q)H(q)}\right]L_{o}(\beta^2q^2)\exp(-\beta^2q^2)dq, 
\label{subeq:102}
\end{equation}
\end{subequations}
\end{widetext}
where
\begin{equation}
	G(q)=2\int^{\infty}_{-\infty}dz_{1}f^{2}_{o}(z_{1})\int^{\infty}_{-\infty}dz_{2}f^{2}_{o}(z_{2})R(q,z_{1}-z_{2}).
\end{equation}

Function $\Pi(q)$ describes the polarization. We take~\cite{ando} 
\begin{equation}
\Pi(q)= \left\{ \begin{array}{ll}
-\frac{2}{q} & \textrm{$(q<2k_{F})$}\\
-\frac{2}{q}\left[1-\sqrt{1-(\frac{2k_{F}}{q})^2}\right] & \textrm{$(q>2k_{F})$}
\end{array} \right.
\end{equation}
where $k_{F}=(2\pi Na^{2}_{B})^{1/2}$ is the Fermi wave vector and $N$ is the density of 2D electrons. Formula (12) is valid for the complete degeneracy of the 2D electron gas at $T\approx 0$ and it neglects the influence of magnetic field on the screening, see Ref.~\cite{raymond}.

Considering a heterojunction GaAs/Ga$_{1-x}$Al$_{x}$As, we deal with a potential well $U(z)$ which takes into account the band offset, the depletion charge and the electron charge in the well. We take the Ando trial function $f_{o}(z)$ for the lowest electric subband~\cite{ando_1}. This function includes the penetration of confined electrons into the Ga$_{1-x}$Al$_{x}$As region due to a finite height of the barrier. This penetration is of importance for impurities located near the interface. We perform the calculations for the Al content $x=0.33$, which corresponds to the barrier height (offset) $V_0=0.257$ eV, and the effective masses  $m^{*}(GaAs)=0.066$ {\itshape m}$_{o}$ and $m^{*}(Ga_{0.67}Al_{0.33}As)=0.073$ {\itshape m}$_{o}$. The dielectric constant $\epsilon=12.9$ is assumed to be the same throughout the entire structure. For the 2D electron density and the depletion density we take typical values of $N=1.36\times10^{11}$ cm$^{-2}$ and $N_{depl}=6\times10^{10}$ cm$^{-2}$, respectively. Functions $R(q,z)$, $H(q)$, and $G(q)$, defined by Eqs.(9) and (11), can be evaluated analytically. However, the corresponding formulas are rather lengthy and they are not shown here. Due to the simple form of the trial function (7), the variational calculations require computation of only one-dimensional integrals ({\itshape cf}. Eqs. (8) and (10)).

We want to compare results for two-electron energies with those of one-electron ones. To this end we calculated also the ground state energy $E_{1e}$ of the one-electron state localized by a negatively charged acceptor. We used the trial function
\begin{equation}
	F_{1e}(\rho)=\frac{1}{\lambda\sqrt{2\pi}}\exp(-\frac{\rho^2}{4\lambda^{2}}).
\end{equation}
where $\lambda$ is a single variational parameter. This function is simpler than that used in Ref.~\cite{kubisa_2} and it is expected to give good results at high magnetic fields. In analogy to the one-electron binding energies of $D^{o}$ and $D^{-}$ centers~\cite{larsen_1}, we define  
\begin{subequations}
\begin{equation}
	E_{r}(1e)= E_{1e}-\gamma
\end{equation}
as the energy required to localize the first electron on a negative acceptor ion and
\begin{equation}
	E_{r}(2e)= E_{2e}-(E_{1e}+\gamma)
\end{equation}
\end{subequations}
as the energy required to add the second electron to the acceptor already occupied by one electron. Here $E_{2e}$ is the energy of the two-electron ground state while $E_{1e}$ is the corresponding energy of the one-electron state. These are evaluated using the trial functions (7) and (14), respectively. The quantity $\gamma$ is equal to the lowest energy of the free electron on the ground Landau level.
\begin{figure} 
\includegraphics[]{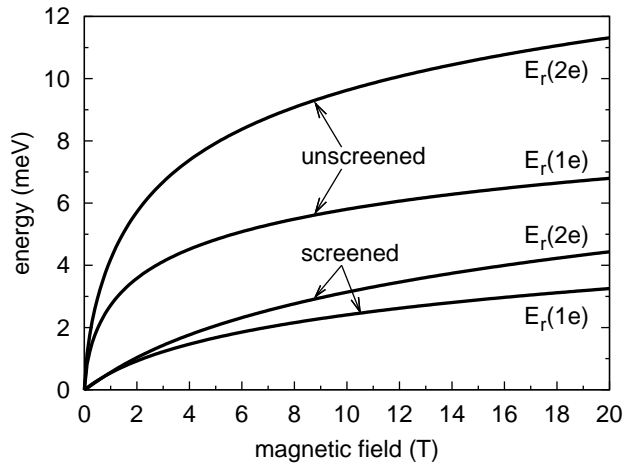}{\centering}
\vspace{1.6cm}
\caption{\label{fig:1}
Calculated one-electron and second-electron repulsive energies of conduction electrons confined to the proximity of a negatively charged acceptor ion by a combined effect of the quantum well and a magnetic field in a GaAs/GaAlAs heterostructure, see text. Width of GaAs layer is $25$ nm, the acceptor layer is $2.5$ nm from the interface with the donor-doped GaAlAs barrier. }
\end{figure}

In Fig. 1 we show the main results of our theory for a GaAs/GaAlAs heterostructure. It is seen that both one-electron and two-electron energies vanish at $B=0$, which reflects the fact that at a vanishing magnetic field there are no localized electron states. One can also see that the screening considerably diminishes the repulsive energies. One should mention that the calculated one-electron energies are noticeably smaller than those obtained in Ref.~\cite{kubisa_2}. The reason is that, as we mentioned above, the presently used trial functions are simpler.

For the unscreened energies there is $E_{r}(2e) \approx 2\cdot E_{r}(1e)$, which can be easily understood: for the first electron the repulsive charge is {\itshape -e} while for the second electron the repulsive charge is 
$-2${\itshape e} (one immobile and one mobile electron). The difference between $E_{r}(1e)$ and $E_{r}(2e)$ becomes smaller with the screening. It is because the Coulomb interaction between two mobile charges is screened more strongly [see Eq. (10a)] than that between a mobile and an immobile charge [see Eq. (10b)]. All in all, it follows from Fig. 1 that the total energy necessary to localize the second conduction electron by an acceptor is higher than that necessary for the first electron. As a consequence, one expects that, in an experiment, one first populates all available acceptor ions with single electrons and only then begins to populate them with second electrons. We show in the experimental part that this is what indeed happens.

\section{EXPERIMENT}
Our magneto-transport experiments were carried out on GaAs/Ga$_{0.73}$Al$_{0.27}$As heterostructures delta-doped in the GaAlAs barrier with Si donors and in the GaAs well with Be acceptors. Preliminary results of these experiments were reported in Ref.~\cite{bisotto} but we show them here for completeness. The experiments were performed on symmetric double-cross samples with the current imposed and the voltage drop measured across the sample. This was done with a dc current source Keithley $220$ and HP $33401$ voltmeters to measure the Hall voltage ($V_{xy}$) and the Shubnikov-de Hass voltage ($V_{xx}$) within a current intensity range $10$ nA - $10$ $\mu$A. In samples not doped with acceptors the quantized Hall breakdown was observed when the current was sufficiently high, in agreement with the observations of other authors. However, for samples doped with acceptors and particularly for the sample $35$A$55$, in which the density of Be atoms {\itshape N}$_a=4{\times}10^{10}$ cm$^{-2}$ was nearly equal to one third of the 2D electron density $N$, well before the breakdown a very strong increase was observed for both $\rho_{xx}$ and $\rho_{xy}$ at high magnetic field (for the filling factor $\nu<1$). This is shown in Fig. 2. 
\begin{figure}
\includegraphics[scale=1.0]{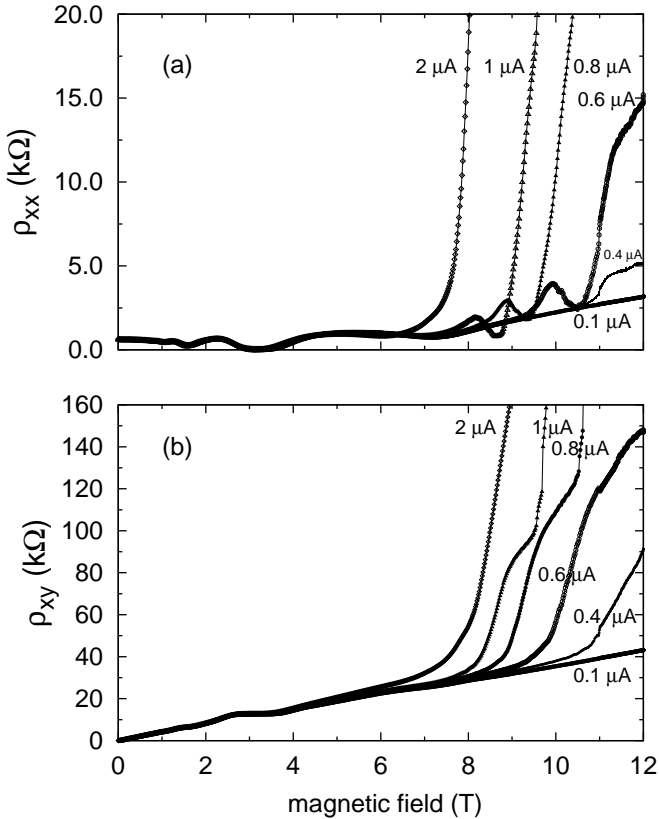}
\vspace{+0.3cm}
\caption{\label{fig:2}
Magneto-transport characteristics of sample 35A55 doped in the well with acceptors, as measured in dc experiments. At higher currents a sharp increase of both $\rho_{xx}$ and $\rho_{xy}$ is observed in the ultra quantum limit of magnetic field $\nu<1$.}
\end{figure}

The large increase of resistivity components is interpreted as a consequence of a strong decrease of 2D electron density $N$ caused by an increasing Hall electric field $F_y$. In our experiments the Hall field is always stronger than the applied driving field $F_x$. It was demonstrated theoretically in Ref.~\cite{bisotto} that a sufficiently high Hall field induces transitions of electrons into empty localized magneto-acceptor states (the boil-off effect). This effect diminishes the density of conducting electrons. The existence of empty acceptor states implies that the Fermi level is below the acceptor level, which occurs in the ultra quantum limit. In the \textit{dc} conditions with a stabilized current, the boil-off process has an avalanche character because the increase of resistivity at a constant current induces a higher driving field. This results in a higher Hall field driving more electrons to the localized acceptor levels which in turn increases the resistivity even further and so on. If experiments are carried in the conditions stabilizing the driving voltage rather than the current, the boil-off effect is still observed, but it does not have an avalanche character, see Ref.~\cite{bisotto}.

The phonon and impurity mechanisms causing electron transitions between the free electron states and the {\itshape A}$^{1-}$ states were described and discussed in Ref.~\cite{bisotto}. Here we will indicate the transfer processes schematically. The free electron energies in crossed electric and magnetic fields are
\begin{equation}
E = \hbar \omega_{c}(n+\frac{1}{2})+eF_{y}y_{o}-\frac{1}{2}m^{*}\frac{F^{2}_{y}}{B^2},
\label{eq:1}
\end{equation}
where $y_{o}=k_{x}(\hbar c/eB)$ is the center of the magnetic motion if one chooses the asymmetric gauge for the magnetic potential: $\textbf{A}=[-By,0,0]$. It is seen that in the presence of electric field the electron energy depends on its position $y_o$ in the sample. In the following we consider the lowest Landau level $0^+$ and omit the shift of all levels given by the last term in Eq. (15). As shown above, the localized {\itshape A}$^{n-}$  states have their energies above the free Landau energies. Figure 3 shows schematically the energy levels across the sample in the model of uniform electric field and the corresponding density of states. 
\begin{figure}
\includegraphics[scale=0.45]{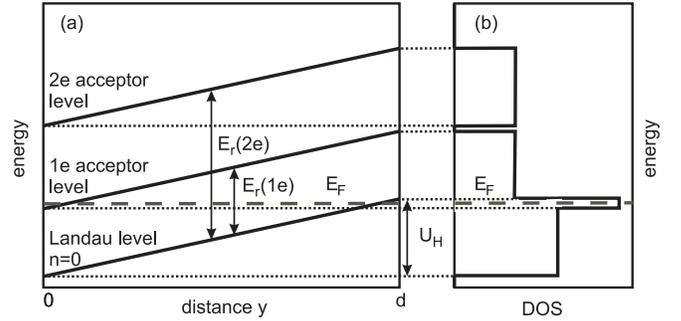}
\vspace{0cm}
\caption{\label{fig:3}  
(a) Free electron and magneto-acceptor energies in the sample in crossed electric and magnetic fields (schematically). (b) Corresponding density of states (DOS). The boil-off process takes place when the Landau level at the right sample edge and the {\itshape A}$^{1-}$ level at the left sample edge begin to overlap.
}
\end{figure}
Measuring $\rho_{xx}$ and $\rho_{xy}$ resistivities it is possible to determine the free electron density $N$ in the sample. Elementary considerations for the transport in a magnetic field lead to the formula, see Ref.~\cite{mansfield}
\begin{equation}
N = \frac{B \rho_{xy}}{e(\rho^{2}_{xy}+\rho^{2}_{xx})}.
\label{eq:16}
\end{equation}
We intend to show the electron density $N(B)$ determined from the data given in Fig. 2 with the use of formula (16) for the acceptor-doped sample. However, in order to appreciate the effects introduced by the acceptors, we show first for comparison in Fig. 4 density $N(B)$ determined by the same procedure for a reference sample not doped intentionally in the well. It is seen that $N(B)$ in the Quantum Hall regime oscillates a little around the value at $B=0$. The linear increases of $N(B)$ correspond to Quantum Hall plateaus. We believe that these small oscillations are due to an electron transfer between the GaAs quantum well and an outside reservoir, as discussed in detail in Ref.~\cite{zawadzki_1}.
\begin{figure}
\includegraphics[scale=0.96]{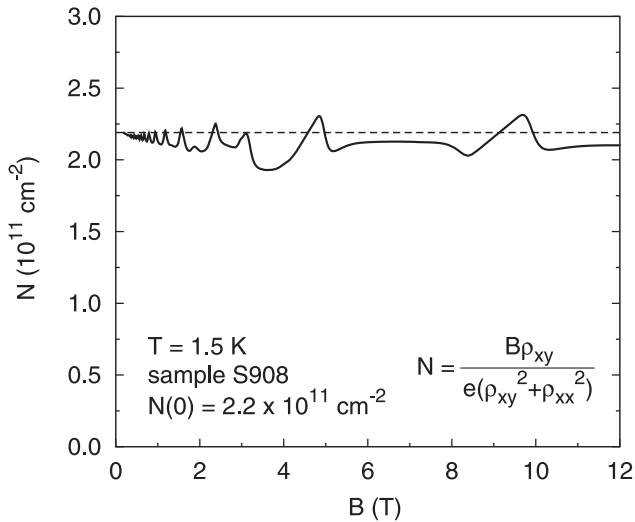}
\vspace{0.0cm}
\caption{\label{fig:4} 
Conduction electron density $N$ versus magnetic field measured for a reference GaAs/GaAlAs heteroctructure not doped intentionally in the well. The density $N(B)$ oscillates slightly around its value $N(0)$.
}
\end{figure}

In contrast, the electron density for the acceptor-doped sample 35A55, as given in Fig. 5, presents a completely different picture. This figure, showing $N(B)$ at different fixed driving currents $I_x$, is our main experimental result. At low magnetic fields, which correspond to weak driving and Hall electric fields, $N(B)$ oscillates around the density value $N(0)$ at $B=0$. The linear increase of $N$ around $B = 3$ T corresponds to the quantum Hall plateau at $\nu=2$. At fields higher than $4$ T the density is higher than $N(0)$ which, we believe, is due to the rain-down effect of electrons falling from the localized {\itshape A}$^{1-}$ states to the conducting free electron Landau states. Finally, for the filling factor $\nu<1$ and sufficiently large currents, a decrease of the density is observed (stronger for higher currents), corresponding to the boil-off of conduction electrons transferred from the free Landau states to the consecutive localized {\itshape A}$^{n-}$ states. One should bear in mind that, as we explained above, higher magnetic fields result in stronger Hall fields, so the boil-off regime corresponds to much higher electric fields than the rain-down regime. Also, higher driving currents result in higher Hall fields, in consequence the strong decrease of electron density due to the boil-off occurs at lower magnetic fields.

\begin{figure}
\includegraphics[scale=0.96]{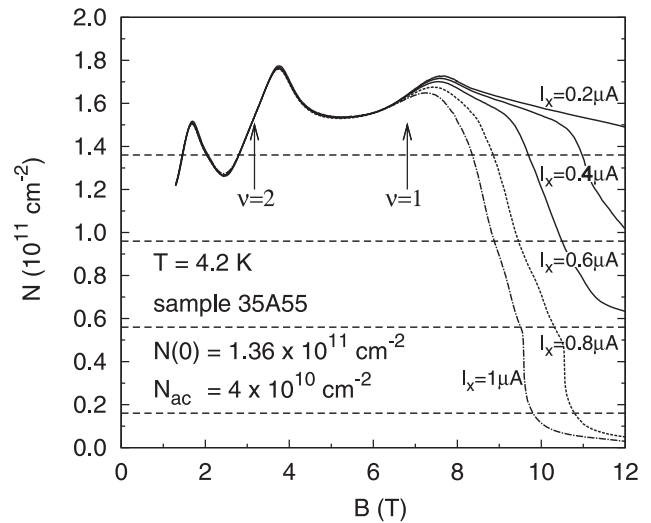} 
\vspace{-0.5cm}
\caption{\label{fig:5} 
Conduction electron density $N$ versus magnetic field at different constant currents, as determined from Fig. 2 and Eq. (16) for the GaAs/GaAlAs sample 35A55 doped in the well with Be acceptors. As the field $B$ increases, the 2D electrons participate consecutively in the quantized Hall effect, the rain-down effect and the boil-off effect, see text.
}
\end{figure}

One can try to estimate experimentally the repulsive energy $E_{r}(ne)$ corresponding to the offset of boil-off for consecutive electrons, if one assumes that the electric field in the sample is homogeneous, see Fig. 3. One can read directly $\rho_{xy}$ from Fig. 2. On the other hand, it follows from Fig. 3 that $U_H=eV_H=eI_x\rho_{xy}=E_{r}(ne)$. For $I_x = 0.2$ $\mu$A  the onset of the decrease of $N$ is at $B = 7.6$ T. One reads in Fig. 2 $\rho_{xy} = 27.7$ k$\Omega$ for the above current and field, which gives $E_{r}(1e)\approx 5$ meV in a reasonable agreement with the theoretical result shown in Fig. 1 for the unscreened regime. However, we do not attach too much importance to this estimation since the Hall electric field in the sample is probably nonhomogeneous, see e.g. Ref.~\cite{weis}.

It is remarkable that all the incidents seen in Fig. 5 on the curves for various currents occur for the density intervals $\Delta N\approx 4\times 10^{10}$ cm$^{-2}$, equal to the number of acceptors. This strongly indicates that the decrease of {\itshape N} as a function of {\itshape B} is related to the acceptors. It is seen that at sufficiently strong magnetic fields the free electron density falls almost to zero. In our interpretation this indicates that almost all electrons are localized by magneto-acceptors. In the sample of our interest the highest electron density is $N(0) \approx 1.36\times 10^{11}$ cm$^{-2}$, while the acceptor density is $N_{a}=4\times 10^{10}$ cm$^{-2}$. This means that at high fields one acceptor localizes roughly four electrons. As mentioned in the Introduction, this result motivated us to undertake the two-electron calculation presented above. Since two localized electrons have a higher repulsive energy than one electron, their localization occurs at a higher magnetic field and the resulting higher Hall electric field. Thus, as both fields increase, the electrons first populate one-electron states of all acceptors, then begin to populate two-electron states, etc. In Fig. 5 one can remark slope discontinuities which occur on the $N(B)$ curves for the currents $0.8$ $\mu$A and 1 $\mu$A around $N= 6\times 10^{10}$ cm$^{-2}$. In our interpretation, these accidents indicate that the electrons begin to populate four-electron states.

\section{DISCUSSION} 
Our work confirms the existence of discrete magneto-acceptor energies in the conduction band above the free-electron Landau levels, related to electron confinement in the vicinity of ionized acceptors by the joint effect of quantum well and external magnetic field. The presented theory of two electrons localized by a charged acceptor supports the experimental findings indicating that an ionized acceptor can localize more than one electron. The presented magneto-transport data on the electron rain-down and boil-off effects concur the evidence provided by photo-magneto-luminescence and cyclotron resonance experiments carried on Be-doped GaAs/GaAlAs heterostructures, see Refs.~\cite{vincente, bonifacie, bisotto}.

On the basis of the presented theory it is qualitatively clear that an ionized acceptor can also localize three or more electrons. The ground states for the third and higher electrons will have higher energies than that for two electrons so that, in order to be populated, they will require higher electric fields in the crossed field configuration. In our arrangement it means that they will be populated at higher magnetic fields. This is what one observes.

We emphasize that the analysis of the electron density behavior $N(B)$, presented in Fig. 5, is based on the classical formula (16) which follows from the Drude formulation of magneto-transport phenomena. This formula should be valid for our purposes since we operate mostly in the non-quantum range of electron behaviour, while the Quantum Hall Effect is seen only for magnetic fields $B<4$ T. The validity of formulas of the type (16) in the quantum range was reviewed in some detail in Ref.~\cite{zawadzki_1}.

It should be mentioned that the increase of conductive electron density, which we attribute to the rain-down effect, was observed also in experiments of other authors, see Refs.~\cite{haug, buth, raymond_1}. In addition, Buth et al.~\cite{buth} in their magneto-transport experiments on acceptor-doped heterostructures found at higher magnetic fields a strong decrease of conducting electron density N, which the authors ascribed to ,,localization of the electrons into the droplet phase''. This observation is similar to ours, but we interpret it above as the localization by acceptor ions. A convincing indication that we deal with the localization of consecutive electrons by acceptors is that one observes "accidents" on the $N(B)$ curves, in intervals equal to the number of acceptors, see Fig. 5.

As mentioned in the Introduction, a positive donor ion with two conduction electrons, i.e. the $D^-$ center, bears some similarity to the negative acceptor ion with two conduction electrons, i.e. the {\itshape A}$^{2-}$ center, considered above. However, physics of the two systems is different. The positive donor ion truly binds the first electron and becomes neutral. The second electron is only bound because the spatial charge distributions of the positive ion and the negative first electron are different. Binding of the second electron is weak and the corresponding binding energy small. As to negative acceptor ion, since the conduction electrons are kept in its proximity only due to the combined effect of the well and magnetic field, the number of such electrons is in principle not limited. The discrete energy of the first electron above the Landau level is determined by the repulsive force of two elementary charges. The energy of the second electron is roughly determined by the repulsive force between the negative ion plus the first electron and the second electron. As a consequence, this energy is larger than that of the first electron, as confirmed by the calculation. The above reasoning can be generalized to further conduction electrons localized by the negative acceptor ion.

Finally, we want to briefly discuss the meaning of words used in the description of {\itshape A}$^{1-}$ and {\itshape A}$^{2-}$ centers, since they were often a source of confusion in the past. In contrast to the donor case, in which the conduction electrons are truly bound by the positive donor ion, the negative acceptor ion does not confine and does not bind the conduction electrons, so that in this case one can not talk about the binding energy. Here the confinement also takes place, but it is a consequence of the quantum well and an external magnetic field. Thus, in the acceptor case one can only talk about the discrete repulsive electron energies above the free-electron Landau levels. The electron (cyclotron) orbits encompass the negative ion and their wave functions are localized in space, as it was explicitly shown in Refs.~\cite{kubisa_1, kubisa_2}. The quantum variational theory shows that one deals with discrete repulsive electron energies for the negative ions located not only in the well but also in the barrier~\cite{kubisa_2}.

As far as future possibilities are concerned, on the theoretical side one can construct a theory for three or more conduction electrons localized by a negative acceptor ion. According to our interpretation, such centers are already observed in the experiments reported above, see Fig. 5. On the experimental side, one should certainly try to investigate other heterostructures than GaAs/GaAlAs doped with other acceptors, for example C atoms. One could also reverse completely the situation by investigating $p$-type heterostructures doped additionally in the well by donors. In general, it seems that the interesting properties of acceptor-doped heterostructures have not been until now sufficiently exploited and applied.

\section{SUMMARY}
We study, both theoretically and experimentally, new discrete quantum states in GaAs/GaAlAs heterostructures created by two conduction electrons localized by a negative acceptor ion the GaAs quantum well. Such a system, which we call the {\itshape A}$^{2-}$ center, is kept together by a combined effect of the well and an external magnetic field $\textbf{\itshape B}$ parallel to the growth direction. The latter provides the Lorentz force keeping the electrons on cyclotron orbits near the negative ion. A variational theory of {\itshape A}$^{2-}$ is presented, showing that the second electron has a considerably higher discrete repulsive energy above the Landau level than the first one. The {\itshape A}$^{n-}$ centers are also studied experimentally with the use of quantum magneto-transport phenomena. Free 2D electron density $N(B)$ is determined in acceptor doped heterostructures and it is shown that in one experimental run one observes the Quantum Hall Effect, the rain-down effect in which electrons fall down from the localized acceptor states to the delocalized Landau states, and the boil-off effect in which the electrons are pushed back by a high Hall electric field from the Landau states to the acceptor states. The repulsive energy for one electron is estimated from the incidents on the $N(B)$ dependence using the homogeneous approximation for the Hall field and shown to be in a reasonable agreement with our variational calculations. At sufficiently high magnetic fields there are almost no conducting electrons left in the sample which provides the evidence that one negative acceptor ion may localize up to four conduction electrons.

\newpage 
%

\end{document}